# Influence of microstructure, temperature and strain ratio on energy approaches for fatigue life prediction

A.L. Gloanec [a], T. Milani [a], G. Hénaff [b]

[a] ENSTA Paristech-UME/MS, 32 boulevard Victor, 75732 Paris Cedex 15, France

[b] LMPM, UMR CNRS 6617, ENSMA, 86961 Futuroscope Chasseneuil, France

**Abstract**

In this paper, two fatigue lifetime prediction models are tested on TiAl intermetallic using results from low-cycle fatigue tests. Both models are based on dissipated energy but one of them considers a hydrostatic pressure correction. This work allows to confirm, on this kind of material, the linear nature, already noticed on silicon molybdenum cast iron, TiNi shape memory alloy and 304L stainless steel, of dissipated energy, corrected or not with hydrostatic pressure, according to the number of cycles to failure. This study also highlights that, firstly, the dissipated energy model is here more adequate to estimate fatigue life and that, secondly, intrinsic parameters like microstructure as well as extrinsic parameters like temperature or strain ratio have an impact on prediction model results.

**Keywords**

TiAl intermetallic, microstructure, temperature, strain ratio, dissipated energy, fatigue life prediction

**Introduction**

During the recent years, alloys based on the intermetallic compound TiAl have attracted a considerable interest as potential competitors to steels and superalloys. Their interesting balance of mechanical properties in the range 600–800°C associated with a low density allows weight saving in some components in jet engines or helicopter turbines. Depending on the processing routes (powder metallurgy, casting, forming, machining, joining) selected to manufacture a given component, a broad range of microstructures (equiaxed γ grains, two-phase nearly lamellar, duplex) can be achieved. It is well established that differences in the final microstructure lead to very different mechanical properties [1-8].

Meanwhile many criteria have been developed to predict the number of cycles to failure. This paper will focus on two criteria based on energy approach. The first one, proposed by Constantinescu et al. [9], is based on the dissipated energy per cycle ($D = \alpha N^{\beta}$). It has been successfully applied on thermomecanical fatigue in silicon molybdenum cast iron. Note that the same criterion has been then tested by Moumni et al. [10] on low-cycle fatigue of shape memory alloys. Amiable et al. [11] identified shortcoming using this criterion. Indeed, only the deviatoric part of the stress tensor is taken into account. Undeniably, it gives a good estimate of the lifetime for multiaxial test, but considering thermal fatigue tests, where stress triaxiality is significant, this criterion does not predict the experimental number of cycles to failure. So, in order to take into account this stress triaxiality effect on lifetime, authors [12] developed a new fatigue criterion. The latter has been established on 304L stainless steel, using thermal fatigue results, introducing a hydrostatic pressure term in the former criterion ($D + c\,P_{max} = \alpha N^{\beta}$).

The originality of the present study rests mainly on the application of both criteria on the same material: a quaternary alloy Ti-48Al-2Cr-2Nb (at.%), using low-cycle fatigue (LCF) test results. The objective is to determine which one is more adequate to assess fatigue lifetime for this intermetallic material. Note that microstructure, temperature and strain ratio effects on the low-cycle fatigue behavior have been reported in previous studies [6,7]. That is why, the impact of these parameters on prediction model results will be also analyzed in this paper.

**Experimental procedure**

Material

The material under study is a quaternary alloy Ti-48Al-2Cr-2Nb (at.%).

The cast alloy is provided by SNECMA MOTEURS. The pieces are initially hipped and then heat-treated. The macrostructure is characterized by two zones. The first one, starting from the surface toward the centre, is made of columnar grains, with the elongation axis oriented along the radial direction, with an average size of about 200–500 µm. The second zone contains equiaxed grains with a grain size of about 50–100 µm. The microstructure in both zones is nearly fully lamellar.

The powder metallurgy (PM) alloy is provided by TURBOMECA. The pieces are simply hipped. The resulting microstructure mainly consists in fine, equiaxed γ-grains, 10-50 µm in size, with $\alpha_2$ phase present along the γ-grain boundaries.

Both microstructures have been detailed elsewhere [6,7,13].

Mechanical testing

LCF tests are conducted on an electro-mechanical closed-loop test system (INSTRON 1362) at various total-strain amplitudes ranging from 0.2% to 0.8% in laboratory air. Since the brittle-to-ductile transition in this alloy occurs at about 700°C, two temperature levels are considered: 25°C and 750°C, below and above the transition temperature, respectively. Specimens are generally fatigued until failure unless otherwise specified. The number of cycles to failure is considered as the fatigue life noted $N_f$. The total-strain amplitude is controlled by means of an INSTRON extensometer, with measured gauge portion of

$15 \pm 0.5$ mm, placed on the gauge of test specimens. The input strain signal shape is triangular. The basic characterization is based on the results of tests conducted with a constant strain rate: $10^{-3}$ s$^{-1}$. Two values of strain ratios ($R_\varepsilon = \varepsilon_{min} / \varepsilon_{max}$) are considered: $R_\varepsilon = -1$ (fully reversed, mean strain equal to zero) and $R_\varepsilon = 0$ (positive mean strain). The cyclic stress-strain (CSS) loops are periodically recorded using an analog plotter. In addition, digital files of CSS loops are acquired by computer for several cycles at close intervals.

**Results and Discussion**

*CSS behavior*

The TiAl intermetallic CSS behavior has been already detailed elsewhere [7,14-16]. The authors reported temperature, microstructure and strain ratio effects on the LCF behavior.

Figure 1 shows temperature effect on the LCF behavior.

At room temperature, the alloy continuously hardens during cycling until failure. This phenomenon is more significant at high total-strain-amplitude and during the first ten cycles. In contrast with room temperature, at 750°C, the stress-amplitude, does not evolve during cycling. At this temperature, the behavior exhibits no hardening and no softening during cycling along the fatigue life, but a cyclic stability of the stress-amplitude at each total-strain-amplitude level investigated. The same temperature dependence the CSS behavior has been reported on the PM alloy [17].

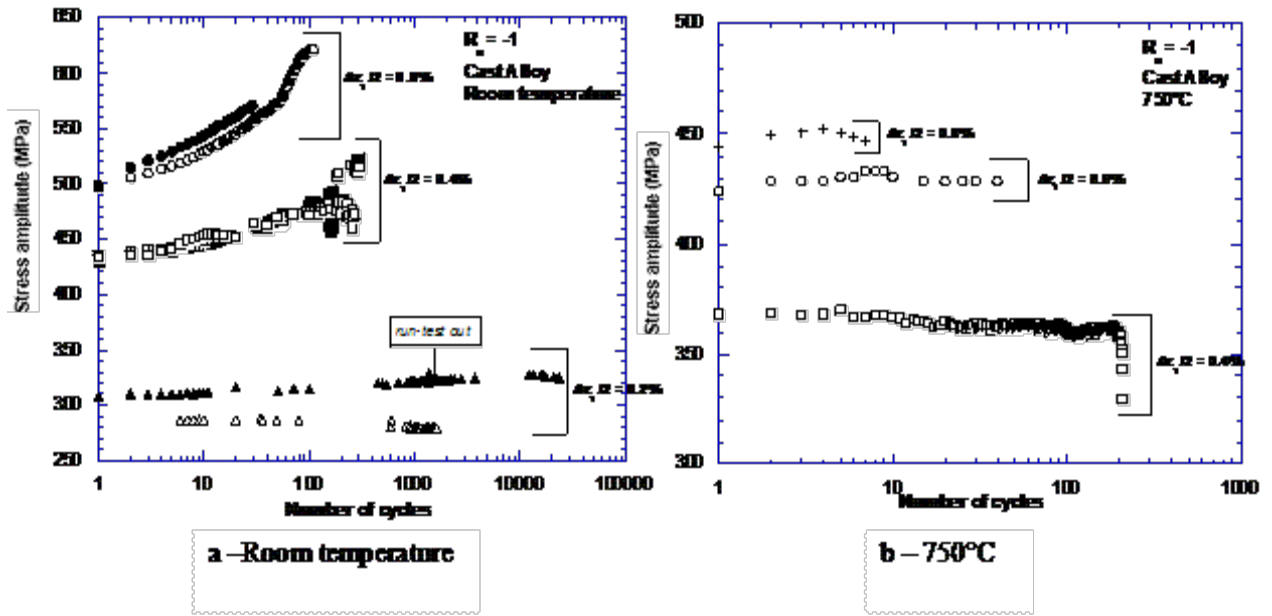

Figure 1: Stress amplitude versus number of cycles for cast alloy at $R_\varepsilon = -1$ at room temperature (a - ) and at 750°C (b - )

Figure 2 represents microstructure effect on the LCF behavior.

Gloanec [17] noted that the CSS behavior of both microstructures is very similar. For high total-strain-amplitude, a significant hardening until failure with a relatively close stress-amplitude is observed. However one has to note that the number of cycles of failure is much shorter for PM alloy. Indeed, for this microstructure, the presence of internal defects (chemical heterogeneity) leads to a premature failure of the test specimen.

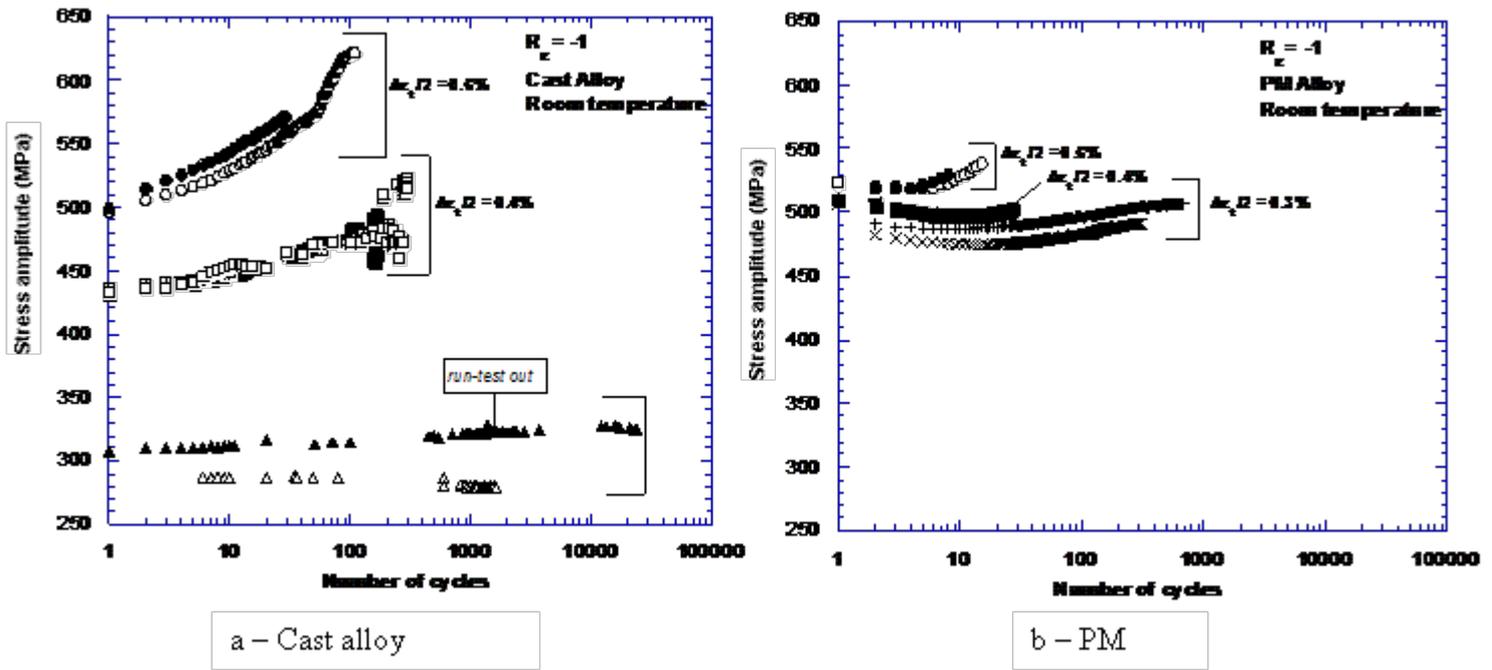

Figure 2: Stress amplitude versus number of cycles for cast alloy (a -) and PM alloy (b -) at $R_\varepsilon$ = -1 and at room temperature

Strain ratio effect on the LCF behavior is reported on Figure 3.

At room temperature, the hardening rate does not seem to be influenced by the strain ratio. The material cyclically hardens regardless the total-strain-amplitude and the strain ratio ($R_\varepsilon$ = -1 or $R_\varepsilon$ = 0). The CSS behavior, for the specimens tested with a null strain ratio, seems to be similar at room temperature and at 750°C, i.e. a slight hardening is observed during the first ten cycles followed by a stability of the strain-amplitude. This slight hardening is not observed for the specimens tested with a negative strain ratio at 750°C.

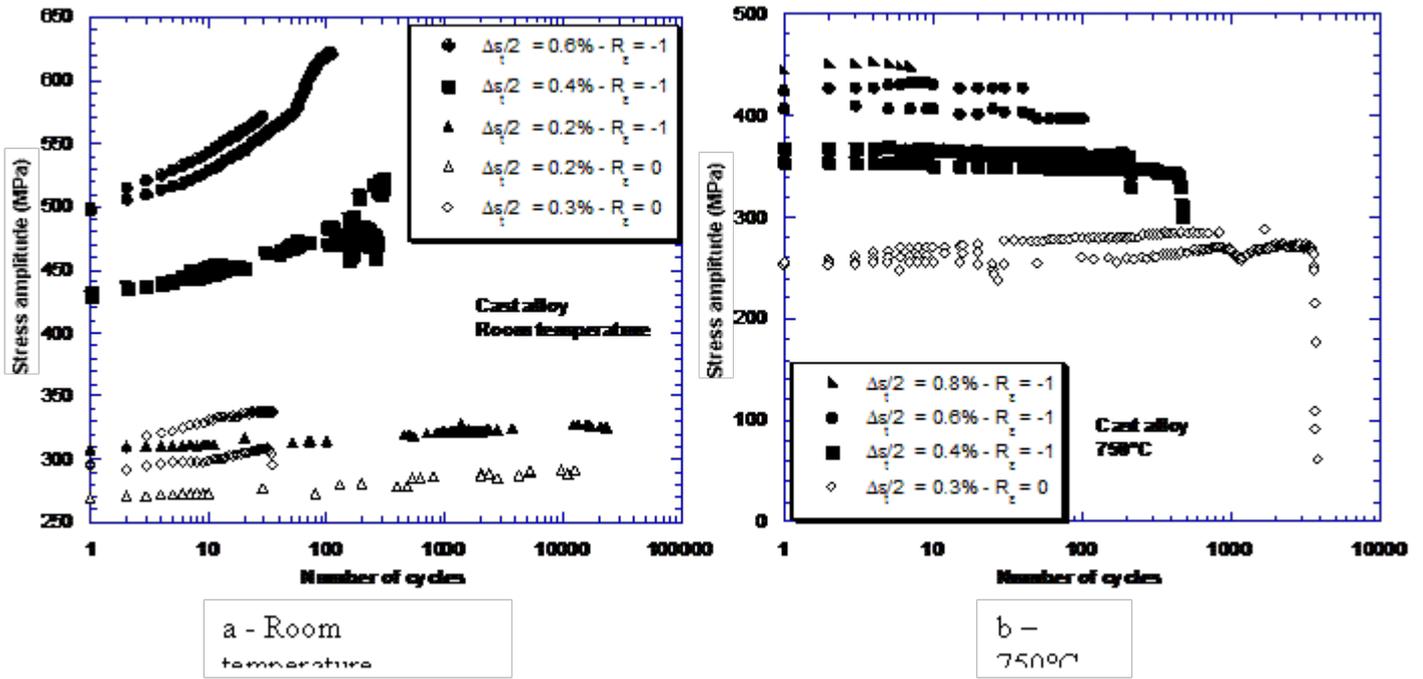

Figure 3: Stress amplitude versus number of cycles for cast alloy at $R_\varepsilon = -1$ and $R_\varepsilon = 0$ at room temperature (a -) and at 750°C (b -)

Let us see here if these same parameters influence both energy approach prediction model results.

*First fatigue criterion: $D = \alpha N^\beta$*

The first criterion tested in this work has been proposed by Constantinescu et al. [9] on silicon molybdenum cast iron, using thermomecanical fatigue test results. This criterion is based on the dissipated energy per cycle and defined as:

$$D = \alpha N^\beta.$$

The dissipated energy, D, is equal to the surface of the hysteresis loop in the strain-stress curve. The damage parameter, D, is calculated for a number of cycles (N) corresponding to a stabilized strain-stress state or for the number of cycles to failure ($N_f$) if the CSS behavior does not present stabilization.

The variations of the dissipated energy according to the number of cycles (N or $N_f$) are given on Figure 4. The linear character of these variations can be easily highlighted for the quaternary alloy Ti-48Al-2Cr-2Nb used in this work. Note that such a phenomenon was already observed by Constantinescu et al. [9] on silicon molybdenum and by Moumni et al. [10] on shape memory alloys. Nevertheless, some differences appear in comparison with literature results. For Constantinescu et al. [9], temperature does not have an influence on the energy value. Here, temperature as well as microstructure have a significant impact on the dissipated energy value. For example, one may observe, on Figure 4-a, for a given temperature and a given number of cycles, a difference, in dissipated energy values, reaching approximately one decade according to microstructure. This phenomenon is less considerable but always present for a given microstructure and a given number of cycles.

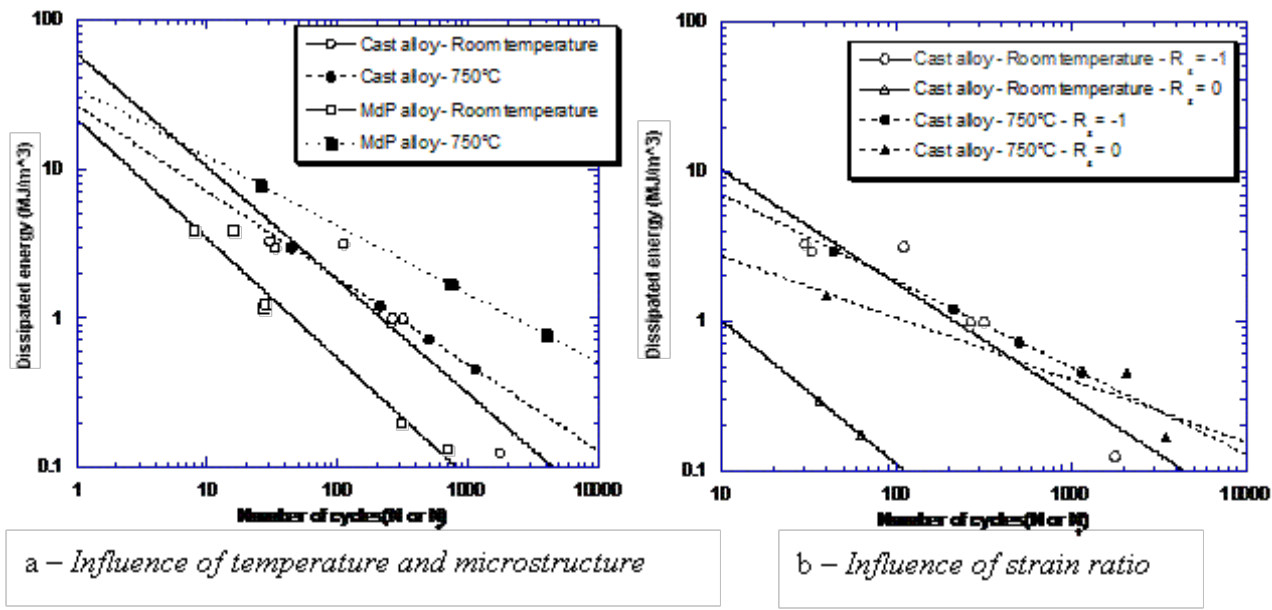

Figure 4: Influence of microstructure, temperature and strain ratio on dissipated energy

LCF experiments carried out by Moumni et al. [10] on shape memory alloys were conducted under stress amplitude control at various stress ratio ($R_\sigma = \sigma_{min} / \sigma_{max}$ = 0, 0.2 and -1). It was shown that in this case, the stress ratio does not have an influence on the dissipated energy value. LCF tests, carried out here on TiAl cast alloy, has been conducted at various total-strain amplitudes with two values of strain ratio ($R_\varepsilon = \varepsilon_{min} / \varepsilon_{max}$ = 0 and -1). Figure 4-b shows a considerable impact of this parameter on dissipated energy values. Here, for a fixed microstructure (lamellar) and a fixed temperature (room temperature), dissipated energy values may vary by as much as one decade for the same cycle number to failure according to strain ratio values. This phenomenon seems to be more important at room temperature than at elevated temperature.

Contrary to Constantinescu et al. [9] or Moumni et al. [10], we noticed that fatigue criterion coefficients, $\alpha$ et $\beta$, are strongly dependent on microstructure, temperature or strain ratio (see Figure 5).

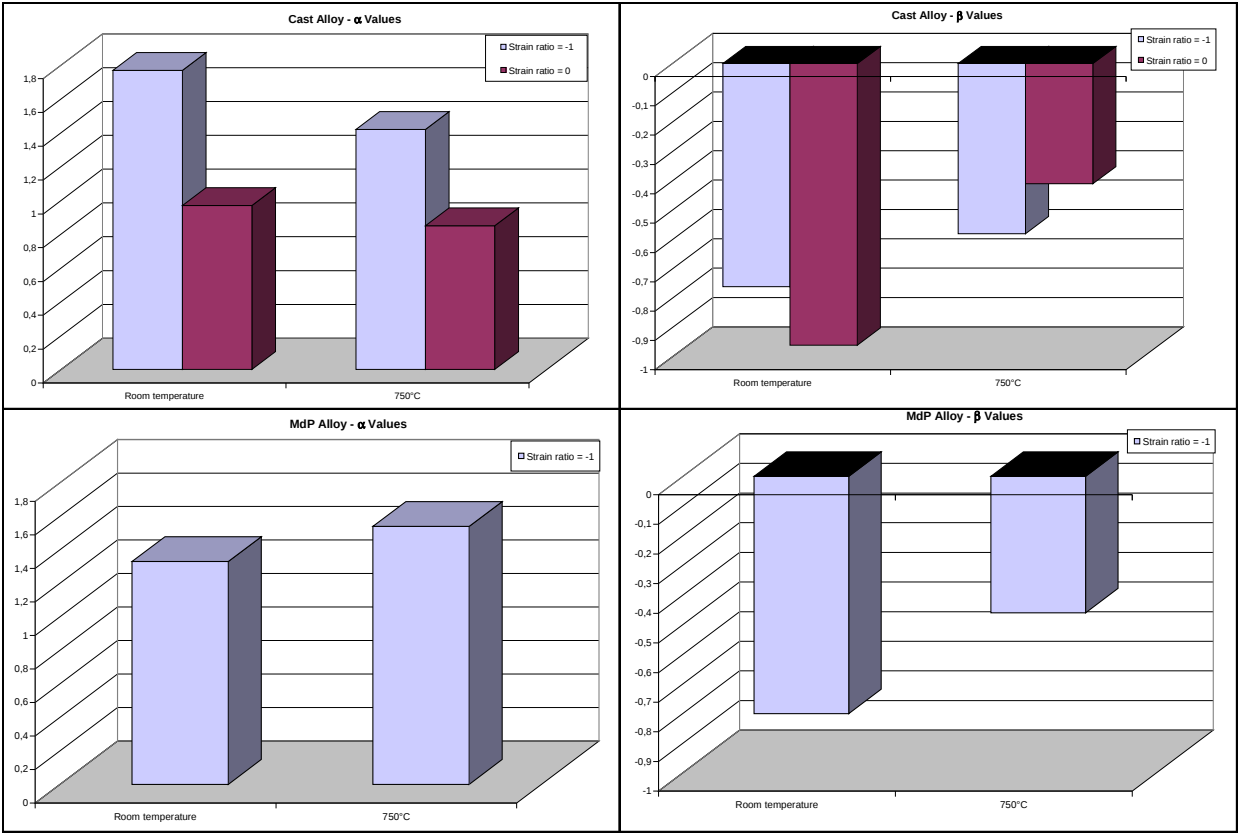

Figure 5: Constantinecu's criterion $\alpha$ / $\beta$ values for both microstructures at various temperatures and strain ratio

The experimental versus predicted lifetime for the LCF tests is presented in log-log plot in Figure 6. On Figure 6-a are reported the results taking into account the influence of strain ratio. In order to compare with the results on shape memory alloys, two values of strain ratio were not dissociated. These results are presented on Figure 6-b.

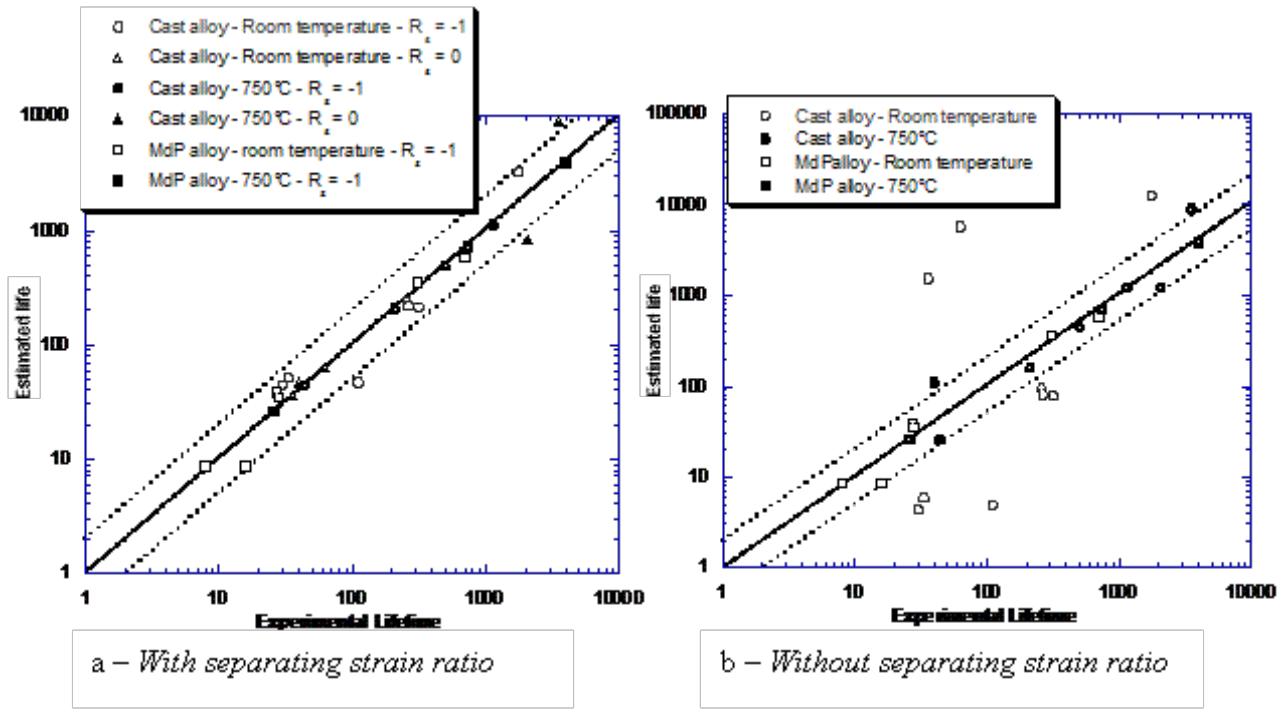

Figure 6: Comparison of Constantinescu's criterion predicted lifetime versus experimental lifetime for various test conditions

The lifetime prediction is in good agreement with the experimental observations if strain ratio values are dissociated (see Figure 6-a). Indeed, all the points lie between in zone of half and twice the lifetime. By not separating strain ratio values, that is to say by following the approach of Moumni et al. [10], this fatigue criterion is less efficient. Undeniably, points are more scattered (see Figure 6-b).

*Second fatigue criterion: $(D + c\, P_{max}) = \alpha\, N^{\beta}$*

The second fatigue criterion tested here has been developed by Amiable et al. [12] on a 304L stainless steel, using results from thermal fatigue experiments. According to these authors, the dissipated energy approach by itself did not allow a good lifetime prediction [11], so he proposed to correct the dissipated energy with a hydrostatic pressure term:

$$(D + c\, P_{max}) = \alpha\, N^{\beta},$$

where $P_{max}$ is the maximal hydrostatic pressure reached during the stabilized cycle (N) or during the cycle of failure ($N_f$) and c an additional material parameter.

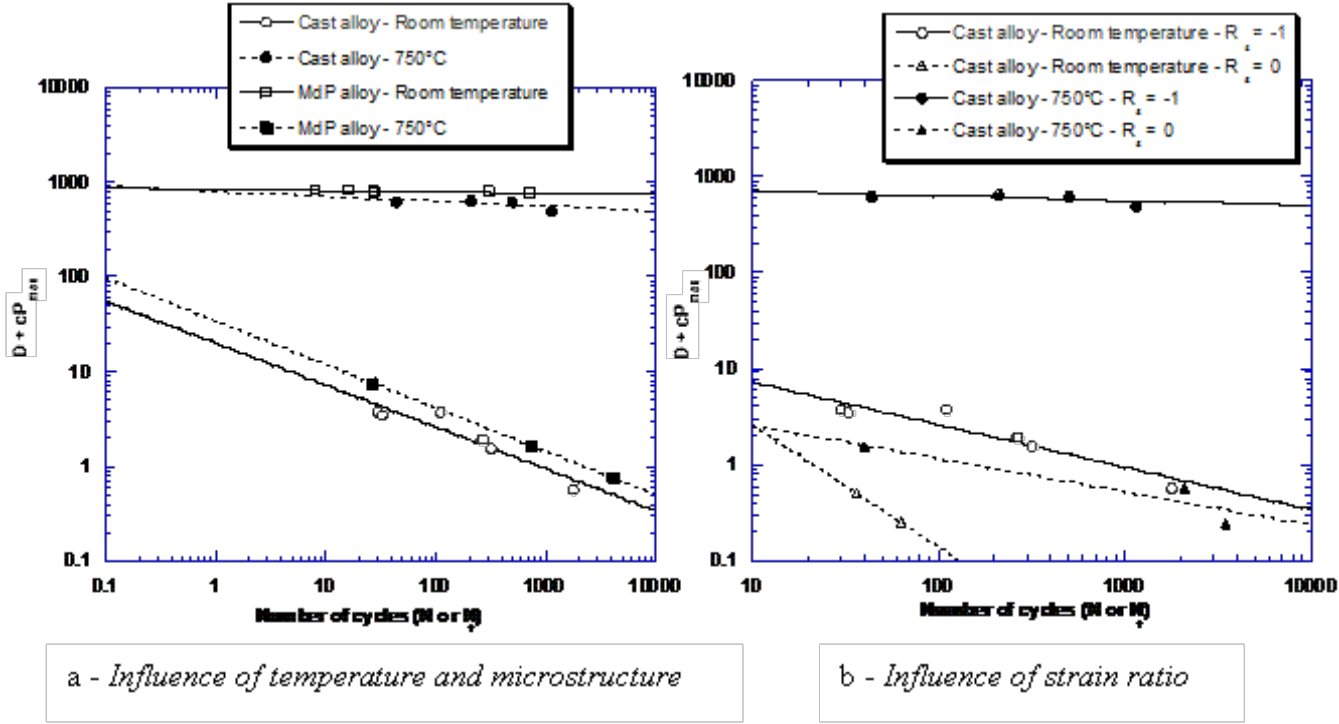

a - *Influence of temperature and microstructure*   b - *Influence of strain ratio*

Figure 7: Influence of temperature, microstructure and strain ratio on the second fatigue criterion values

As reported on Figure 7, the variations of the criterion left-hand term according to the number of cycles show a linear nature for quaternary alloy Ti-48Al-2Cr-2Nb. This phenomenon is always present regardless of microstructure, temperature (see Figure 7-a) or strain ratio (see Figure 7-b). However, an influence of temperature, microstructure or strain ratio is still observed on Amiable's criterion parameters $\alpha$ and $\beta$ (see Figure 8).

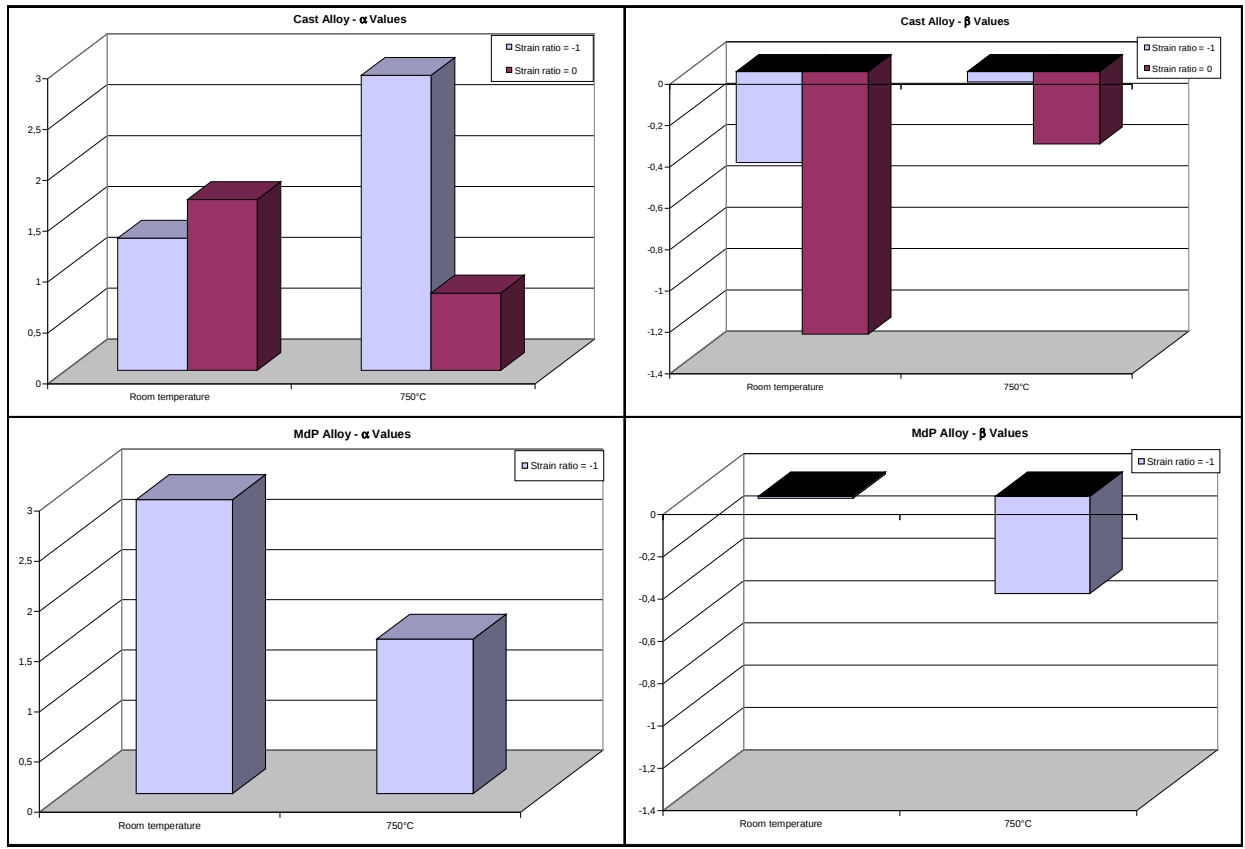

Figure 8: Amiable's criterion $\alpha / \beta$ values for both microstructures at various temperatures and strain ratio

In order to examine the validity of the second fatigue criterion on this material, let us consider a graph giving estimated life versus experimental lifetime (see Figure 9). As for the first criterion used previously, two cases have been taken into account: with separating strain ratio (see Figure 9-a) and without separating strain ratio (see Figure 9-b).

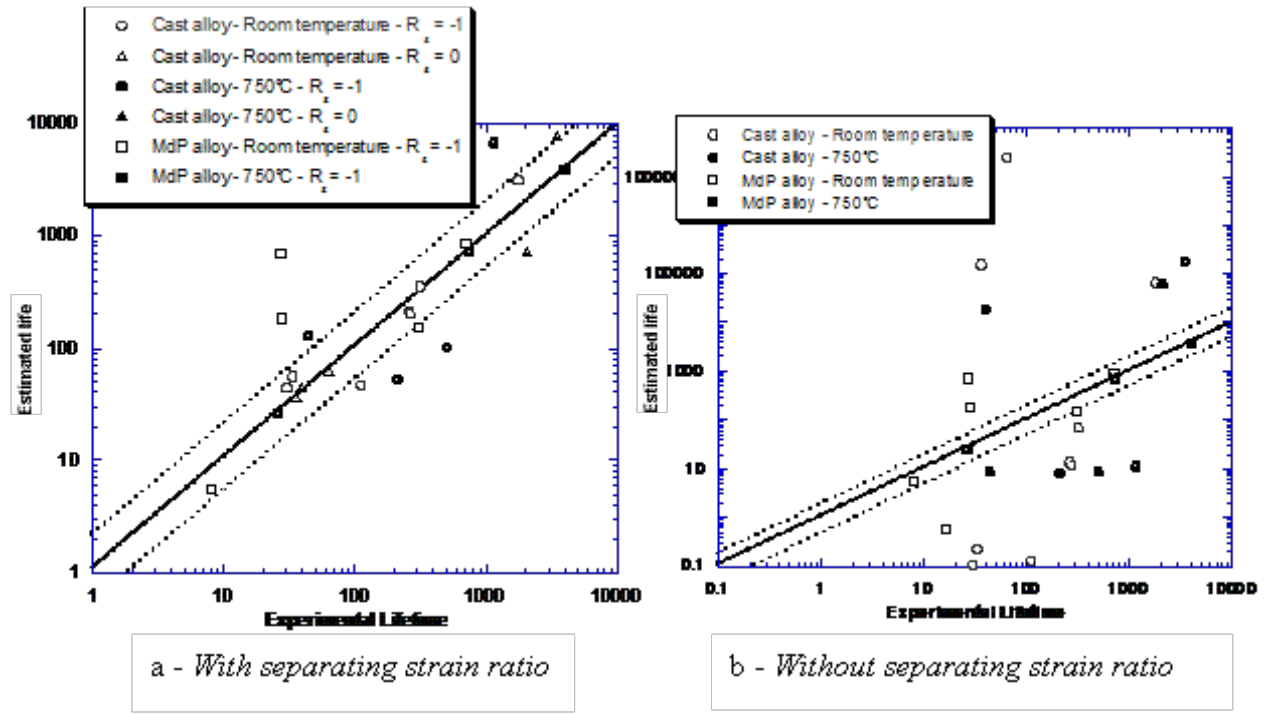

Figure 9: Comparison of Amiable's criterion predicted lifetime versus experimental lifetime for various test conditions

Whatever the case, strain ratio values being dissociated or not, points are scattered. In both cases, many points are apart from the zone corresponding to half and twice the experimental lifetime. Nevertheless, note that scatter is less considerable if strain ratio values are dissociated, just like for the first criterion.

**Conclusion**

The aim of this study was to apply two fatigue criteria on a quaternary alloy Ti-48Al-2Cr-2Nb (at.%) using results from low-cycle fatigue tests. The first one is simply based on dissipated energy ($D = \alpha N^{\beta}$) and the second one on dissipated energy corrected with a hydrostatic pressure term ($D + c\, P_{max} = \alpha N^{\beta}$).

Whatever the fatigue criteria is, a linear nature of the variations of the energy parameters according to the cycle number may be pointed out. Nevertheless this phenomenon is very dependant on microstructure, temperature or strain ratio.

Moreover, for both criteria, strain ratio values have to be taken into account separately in order to have a good agreement between lifetime prediction and experimental observations.

Lastly, a major result consists in finding that, for the quaternary alloy studied here with LCF tests, it is the fatigue criterion just based on dissipated energy ($D = \alpha\ N^{\beta}$) which gives the better lifetime prediction.

**Acknowledgements**

The authors would like to acknowledge the material supply from SNECMA MOTEURS (in particular Marjolaine GRANGE) and TURBOMECA (in particular Philippe BELAYGUE). The results, used for this study, are stemmed from an experimental campaign carried out six years ago. This experimental part was conducted within the framework of a national project (CPR " intermetallic") in collaboration Laboratoire de Thermodynamique et Physico-Chimie Métallurgiques (Grenoble), Centre d'Etudes de Chimie Métallurgique (Vitry), Laboratoire de Science et Génie des Matériaux Métalliques (Nancy), Groupe de Métallurgie Physique (Rouen), Centre d'Elaboration de Matériaux et d'Etudes Structurales (Toulouse), Laboratoire de Mécanique des Solides (Palaiseau), Laboratoire de Mécanique et Physique des Matériaux (Poitiers), the societies SNECMA MOTEURS and TURBOMECA, with the support of CNRS (Centre National de la Recherche Scientifique) and DGA (Délégation Générale de l'Armement).

**References**

[1] Huang S.C. and Chesnutt J.C., Gamma TiAl and its alloys. Serie Gamma TiAl and its alloys, 1994, Vol. 2, pp. 73-90.

[2] Kim Y.W., Intermetallic alloys based on gamma titanium aluminide. Journals of Metals, 1989, Vol. 41, n°7, pp. 24-30.

[3] Appel F. and Wagner R., Microstructure and deformation of two-phase gamma-titanium aluminides. Materials Science and Engeneering A. Reports: A Review Journal, 1998, Vol. R22, pp. 187-268.

[4] Kim Y.W., Microstructural evolution and mechanical properties of a forged gamma titanium aluminide alloy. Acta Materialia, 1992, Vol. 40, n°6, pp. 1121-1134.

[5] Kim Y.W. and Dimiduck D.M., Progress in the understanding of gamma titanium aluminides. Journals of Metals, 1991, Vol. 43, n°8, pp. 40-47.

[6] Gloanec A.L., Henaff G., Bertheau D., Jouiad M., Grange M. and Belaygue P., Fatigue properties of a Ti-48Al-2Cr-2Nb alloy produced by casting and powder metallurgy. Third International Symposium on Gamma Titanium Aluminide (ISGTA III), 2003, Vol. pp. 485.

[7] Gloanec A.L., Henaff G., Bertheau D. and Gadaud P., Influence of temperature, strain ratio and strain rate on the cyclic stress-strain behaviour of a gamma-titanium-aluminide alloy. Matériaux et techniques, 2004, Vol. 1-2, pp. 77-84.

[8] Gloanec A.L., Henaff G., Bertheau D., Belaygue P. and Grange G., Fatigue crack growth behaviour of a gamma-titanium-aluminide alloy prepared by casting and powder metallurgy. Scripta Materialia, 2003, Vol. 49, pp. 825-830.

[9] Constantinescu A., Charkaluk E., Lederer G. and Verger L., A computational approach to thermomechanical fatigue. International Journal of Fatigue, 2004, Vol. 26, pp. 805-818.

[10] Moumni Z., Herpen A. Van and Riberty P., Fatigue analysis of shape memory alloys: energy approach. Smart Materials and Structures, 2005, Vol. 14, pp. S287-S292.

[11] Amiable S., Prédiction de durée de vie de structures sous chargement de fatigue thermique. Thesis of Versailles University, 2006.

[12] Amiable S., Chapuliot S., Constantinescu A. and Fissolo A., A comparison of lifetime prediction methods for a thermal fatigue experiment. International Journal of Fatigue, 2006, Vol. 28, pp. 692-706.

[13] Lagattu F., Gloanec A.L., Hénaff G. and Brillaud J., Etude du rôle de la microstructure sur la résistance à la fissuration par fatigue des alliages TiAl à l'aide de la technique de corrélation d'images numériques de mouchetis. Mécanique et Industrie, 2005, Vol. 6, pp. 499-507.

[14] Henaff G. and Gloanec A.L., Fatigue properties of TiAl alloys. Intermetallics, 2005, Vol. 13, pp. 543-558.

[15] Gloanec A.L., Henaff G., Jouiad M., Bertheau D., Belaygue P. and Grange M., Cyclic deformation mechanisms in a gamma titanium aluminide alloy at room temperature. Scripta Materialia, 2005, Vol. 52, pp. 107-111.

[16] Gloanec A.L., Jouiad M., Grange M. and Henaff G., Cyclic deformation mechanisms in a cast gamma titanium aluminide alloy. Materials Science and Engineering A, 2005, Vol. 400-401, pp. 409-412.

[17] Gloanec A.L., Mécanismes gouvernant le comportemenet cyclique et la résistance à la fissuration des alliages TiAl. Thesis of Poitiers, ENSMA, 2003.

**Figure Captions**